\documentclass[twocolumn,floatfix,superscriptaddress]{revtex4-2}
\pdfoutput=1
\usepackage[utf8]{inputenc}
\usepackage[english]{babel}
\usepackage[T1]{fontenc}
\usepackage{amsmath}
\usepackage{hyperref}
\usepackage[normalem]{ulem}

\usepackage{graphicx}
\usepackage{amsmath,mathtools}
\usepackage{amssymb}
\usepackage{bm}
\usepackage{color}
\usepackage{bbold}
\usepackage{mathtools}

\DeclarePairedDelimiterX\braket[2]{\langle}{\rangle}{#1\,\delimsize\vert\,\mathopen{}#2}
\usepackage{tikz}
\usepackage{quantikz}
\usepackage{lipsum}

\begin{document}

\title{Dynamic thermalization on noisy quantum hardware}

\author{H. Perrin}
\affiliation{University of Strasbourg and CNRS, CESQ and ISIS (UMR 7006), aQCess, 67000 Strasbourg, France}
\affiliation{TKM, Karlsruhe Institute of Technology, 76131 Karlsruhe,
Germany}
\author{T. Scoquart}
\affiliation{TKM, Karlsruhe Institute of Technology, 76131 Karlsruhe,
Germany}
\affiliation{IQMT, Karlsruhe Institute of Technology, 76131 Karlsruhe,
Germany}
\author{A. I. Pavlov}
\email{andrei.pavlov@kit.edu}
\affiliation{IQMT, Karlsruhe Institute of Technology, 76131 Karlsruhe,
Germany}
\author{N. V. Gnezdilov}
\email{nikolay.gnezdilov@dartmouth.edu}
\affiliation{Department of Physics and Astronomy, Dartmouth College, Hanover, NH 03755, USA}

\begin{abstract}
Emulating thermal observables on a digital quantum computer is essential for quantum simulation of many-body physics. However, thermalization typically requires a large system size due to incorporating a thermal bath, whilst limited resources of near-term digital quantum processors allow for simulating relatively small systems. We show that thermal observables and fluctuations may be obtained for a small closed system without a thermal bath. Thermal observables occur upon classically averaging quantum mechanical observables over randomized variants of their time evolution that run independently on a digital quantum processor. Using an IBM quantum computer, we experimentally find thermal occupation probabilities with finite positive and negative temperatures defined by the initial state's energy. Averaging over random evolutions facilitates error mitigation, with the noise contributing to the temperature in the simulated observables. This result fosters probing the dynamical emergence of equilibrium properties of matter at finite temperatures on noisy intermediate-scale quantum hardware.
\end{abstract}

\maketitle

{\bf Introduction.}
The notions of temperature and statistics in isolated quantum systems are based on the mean and variance of a physical observable over repeated experiments \cite{Mueller2010, Sanner2010}. Microscopically, the emergence of statistical mechanics from the quantum description of a closed system is understood via thermalization. Thermalization implies that the quantum mechanical expectation value of the observable dynamically relaxes to the thermal one \cite{Deutsch1991, Srednicki1994, Rigol2008, DAlessio2016, Ueda2020Quantum}. 
Achieving thermalization requires the system to equilibrate: the expectation value of an observable $\cal O$ is stationary,
\begin{equation}
    \langle {\cal O} \rangle_{\rm eq} = \overline{\langle {\cal O} \rangle} = {\rm Tr} \, \overline{\rho} \, {\cal O}, \label{equilibration}
\end{equation}
where $\overline{\rho}$ is the time-averaged diagonal density matrix and the trace is taken over the entire system.
In turn, the equilibrium expectation value (\ref{equilibration}) is thermal if it agrees with the one evaluated with the thermal density matrix $\rho_{\rm th}$ describing a Gibbs state,
\begin{equation}
    {\rm Tr} \, \overline{\rho} \, {\cal O} = {\rm Tr} \, \rho_{\rm th} {\cal O}  = {\cal O}_{\rm th}. \label{thermalization}
\end{equation}

Relaxation after a quench is one way thermal observables occur in a closed quantum system.
The quench induces an interaction between the system's subparts and initiates the dynamics. As a result, the system equilibrates (\ref{equilibration}) and thermalizes (\ref{thermalization}) unless it possesses extra constraints that, e.g. result in integrability or lead to many-body localization \cite{Abanin2011coll}.

In the present research, we demonstrate a mechanism for the occurrence of thermal observables and fluctuations in the course of post-quench dynamics, not relying on the presence of a thermal bath that ensures thermalization if the system is rather small. For an isolated system of several qubits (either interacting or non-interacting), we consider a quench protocol that turns on an all-to-all interaction with coupling constants sampled from a complex Gaussian distribution with zero mean and finite variance. 
The protocol is repeated multiple times with different sets of couplings. We find the equilibrium value of an observable dynamically arising from averaging over the realizations of the quench protocol:
\begin{equation}
    \langle {\cal O} \rangle_{\rm eq} = \langle\langle {\cal O}(t>t_{\rm eq}) \rangle\rangle. \label{equilibration_2}
\end{equation}
The double brackets denote the quantum mechanical mean averaged over realizations, and $t_{\rm eq}$ is the time after which the observable becomes stationary. 
The observable equilibrates to the thermal value, 
\begin{equation}
     \langle\langle {\cal O}(t>t_{\rm eq}) \rangle\rangle =  {\cal O}_{\rm th}(T(\varepsilon_{\rm in})), \label{thermalization_2}
\end{equation}
where the temperature $T$ appears due to the quenched interaction and depends on the energy of the initial state $\varepsilon_{\rm in}$. The system thermalizes to a positive or negative temperature state as it has a bounded spectrum and no kinetic energy \cite{Landau1980}. The average thermal observables emerge irrespective of whether the pre-quenched Hamiltonian of the system is integrable or not. 
Below, we demonstrate this approach leads to thermal observables for the pre-quenched system without an external thermal bath.

We experimentally examine our thermalization mechanism (\ref{equilibration_2},\ref{thermalization_2}) in a simulation on an IBM Quantum computer (IBMQ) for a system of four qubits. Upon equilibration, we read out the occupation probabilities of the many-body energy states of the system described by the pre-quench Hamiltonian. For the system's initial state given by the ground state of the qubits, the average occupation probabilities relax to the Gibbs distribution. Starting from the highest excited state leads to the inversely populated state ($T<0$). As a result, both the observables and their fluctuations evaluated from the noisy quantum computer's error-mitigated output display thermal behavior with positive or negative absolute temperatures. 

Being implemented on noisy intermediate-scale quantum (NISQ) hardware, our approach allows probing thermal observables in finite-size quantum systems in a highly controlled environment and may be utilized among nature-inspired quantum algorithms for thermal state preparation \cite{Chen2023B, Shtanko2023}.
Considerable experimental effort has been spent on sampling observables agreeing with the Gibbs state prediction on quantum computers. Analog or hybrid analog-digital devices allow simulating finite-temperature transitions for systems of $\sim 10$-$70$ qubits \cite{Schuckert2025, andersen2024thermalizationcriticalityanalogdigitalquantum}, but with the trade-off of having no precise control over the target temperature and a lack of universality compared to their purely digital (gate-based) counterparts.
In turn, some recent experiments on digital quantum computers are based on an imaginary time evolution algorithm \cite{Motta2019, Sun2021Quantum} or an approach using thermofield double states \cite{Zhu2020, Sagastizabal2021}, while another incorporates an infinite thermal bath using ancilla qubits \cite{Shtanko2023}. These experiments utilize ancilla qubits and achieve thermalization for relatively small systems of $2$ to $4$ qubits.
Quantum-classical algorithms combining quantum simulations with classical data post-processing, matching quantum output with results expected from classical simulations, allow pushing this limit for larger systems of $\sim 8$ qubits \cite{Tazhigulov2022}.  Here, we also address thermalization in a small system on a digital quantum computer.
Typically, one assesses thermalization from the agreement of the observables measured at the end of the simulation with the Gibbs state prediction. We rely on the post-quench evolution and the averaging of observables over realizations of the randomized auxiliary interaction. Our protocol is inherently dynamical, enabling studying the formation of thermal observables in real time. Furthermore, it relies on the classical averaging of quantum observables corresponding to an individual realization of the quench protocol, unloading precious resources of the quantum computer. Advantageously, it does not require long-time evolution or increasing the system size due to ancilla qubits aiming to simulate a thermal bath. 

The paper is organized as follows. First, we define the quench protocol, introducing the auxiliary random interaction that governs post-quench evolution. Then, we derive thermal energies and occupation probabilities from the dynamic observables averaged over realizations of the auxiliary interaction for a system describing four non-interacting qubits before the quench. We demonstrate that the average total energy conservation respected in the evolution defines the temperature in the equilibrated observables, whilst the choice of the initial state defines the temperature's sign. We use eigenstates of the pre-quenched Hamiltonian as the initial states. Second, we repeat the protocol for a more general case, where we choose a non-integrable Ising model as our pre-quenched Hamiltonian for eight qubits and begin with an arbitrary initial state. Despite the absence of a thermal bath, averaging the observables over random instances of the auxiliary interaction, we report that the energy fluctuations-to-energy ratio for the Ising Hamiltonian equilibrates to a value predicted by temperature determined by the conservation of total energy, reminiscent of fluctuation-dissipation theorem. Third, we run our thermalization protocol on a quantum computer for a system of four qubits, extracting thermal occupations, energies, and energy fluctuations corresponding to finite positive and negative temperatures. Each realization of the auxiliary interaction runs independently on the quantum computer while averaging is done classically upon readout. Implementing error-mitigation, we find that the hardware noise partly contributes to the temperature in the average simulated observables.

{\bf The model.}
Quenching a closed system is a natural way of probing its equilibration dynamics and subsequent thermalization \cite{Polkovnikov2011coll}. For example, thermalization may occur via a non-linear classical mode and the development of turbulence, leading to eventual energy equipartition \cite{Frahm2023}, or due to switching on a highly non-local interaction in a quantum system resulting in its equilibration to an infinite temperature state \cite{Rossini2020, Bandyopadhyay2023}. Taking this dynamical perspective, we organize our quench protocol as follows.

We prepare $N$ non-degenerate qubits in an eigenstate of $H_0= \sum_{j=1}^N (\omega_j/2) \, \sigma^y_j$: $H_0|\psi_{\rm in}\rangle = \varepsilon_{\rm in} |\psi_{\rm in}\rangle$. Here $\sigma_j^y$ is the second Pauli matrix acting on the $j$th qubit, $\omega_j$ is the $j$th qubit's frequency, $|\psi_{\rm in}\rangle$ is the initial state of the system, and $\varepsilon_{\rm in}$ is the corresponding eigenenergy. At time $t=0$, we abruptly turn on a global quench that couples the qubits all-to-all and initiates the dynamics. The initial state $|\psi_{\rm in}\rangle$ starts evolving with the Hamiltonian
\begin{equation}
    H = H_0 + {\cal V}, \label{H}
\end{equation}
where 
\begin{equation}
    {\cal V} = -\sum_{j_1>j_2}^N J_{j_1 j_2} \sigma_{j_1}^+ \prod^{j_1-1}_{k=j_2+1} \! \sigma_k^z \, \sigma_{j_2}^- + h.c. \label{V}
\end{equation}
describes the interaction between the qubits with the coupling constants $J_{j_1 j_2}$ drawn from the Gaussian unitary ensemble with zero mean $\langle J_{j_1 j_2} \rangle = 0$ and finite variance $\langle |J_{j_1 j_2}|^2 \rangle = J^2/N$. 
Accordingly, one may understand the interaction term (\ref{V}) as the Hamiltonian of random non-interacting fermions \cite{Magan2016Random} projected onto the qubits' basis via the Jordan-Wigner transform. While both $H_0$ and ${\cal V}$ are integrable, their combination is not. Indeed, when ${\cal V}$ is combined with $H_0$ in the full Hamiltonian ({\ref{H}}), the latter is chaotic, displaying Wigner-Dyson level statistics \cite{Mehta1991Random} as discussed in more detail in Appendix \ref{app:RMT}. As such, the Hamiltonian ({\ref{H}}) describes an interacting theory in fermionic or qubit representation, giving rise to the quantum many-body dynamics we address below. 

We repeat the time evolution of the initial state $|\psi_{\rm in}\rangle$ with the Hamiltonian (\ref{H}) for different sets of random couplings and compute quantum mechanical mean values averaged over realizations of the quench protocol, aiming to check the equilibration (\ref{equilibration_2}) and thermalization (\ref{thermalization_2}) conditions. 

Once the interaction ${\cal V}$ is on, the evolving state $|\psi (t) \rangle = e^{-it(H_0+{\cal V})} |\psi_{\rm in}\rangle$ is no longer an eigenstate of the initial Hamiltonian $H_0$ whose expectation value $\langle H_0 (t)\rangle = \langle \psi(t) | H_0 | \psi(t)\rangle$ becomes dynamic. To study the dynamics, we write $H_0 = \sum_l \varepsilon_l |l\rangle \langle l|$ in the basis of its $L=2^N$ many-body energy eigenstates labeled by the index $l$, and average the expectation value $\langle H_0 (t)\rangle$ over the protocol's realizations. Thereby, we have
\begin{equation}
    \langle\langle H_0(t) \rangle\rangle  = \sum_{l=0}^{L-1} \varepsilon_l \langle\langle n_l(t) \rangle\rangle, \label{H0_av}
\end{equation}
where 
\begin{equation}
    \langle \langle n_l(t) \rangle \rangle = \langle \langle l| e^{-i t (H_0 +{\cal V})} |\psi_{\rm in}\rangle\langle \psi_{\rm in} | e^{i t (H_0 +{\cal V})} | l \rangle \rangle, \label{n_av} 
\end{equation} 
is the average occupation probability of the many-body state with energy $\varepsilon_l$ for $t>0$. Throughout the evolution, the total energy of the system is conserved on average, $\langle\langle H(t) \rangle\rangle = \langle\langle \psi_{\rm in}| H | \psi_{\rm in} \rangle\rangle= \varepsilon_{\rm in}$, due to the zero mean of the Gaussian in Eq.~(\ref{V}).

{\bf Exact dynamics.}
Anticipating our quantum simulation experiment, we choose $N\!=\!4$ and run the complete protocol numerically for different initial states $|\psi_{\rm in}\rangle \!=\! |\!\downarrow\downarrow \downarrow \downarrow \rangle$,  $|\!\uparrow\downarrow \downarrow \downarrow\rangle$, $\ldots$, $|\!\uparrow\uparrow\uparrow \uparrow\rangle$, all having distinct initial energies, where the down arrow denotes a qubit in the ground state of $\sigma^y_j$ and the up arrow marks the excited state of a qubit. 

\begin{figure*}[t!!]
\center
\includegraphics[width=1.\linewidth]{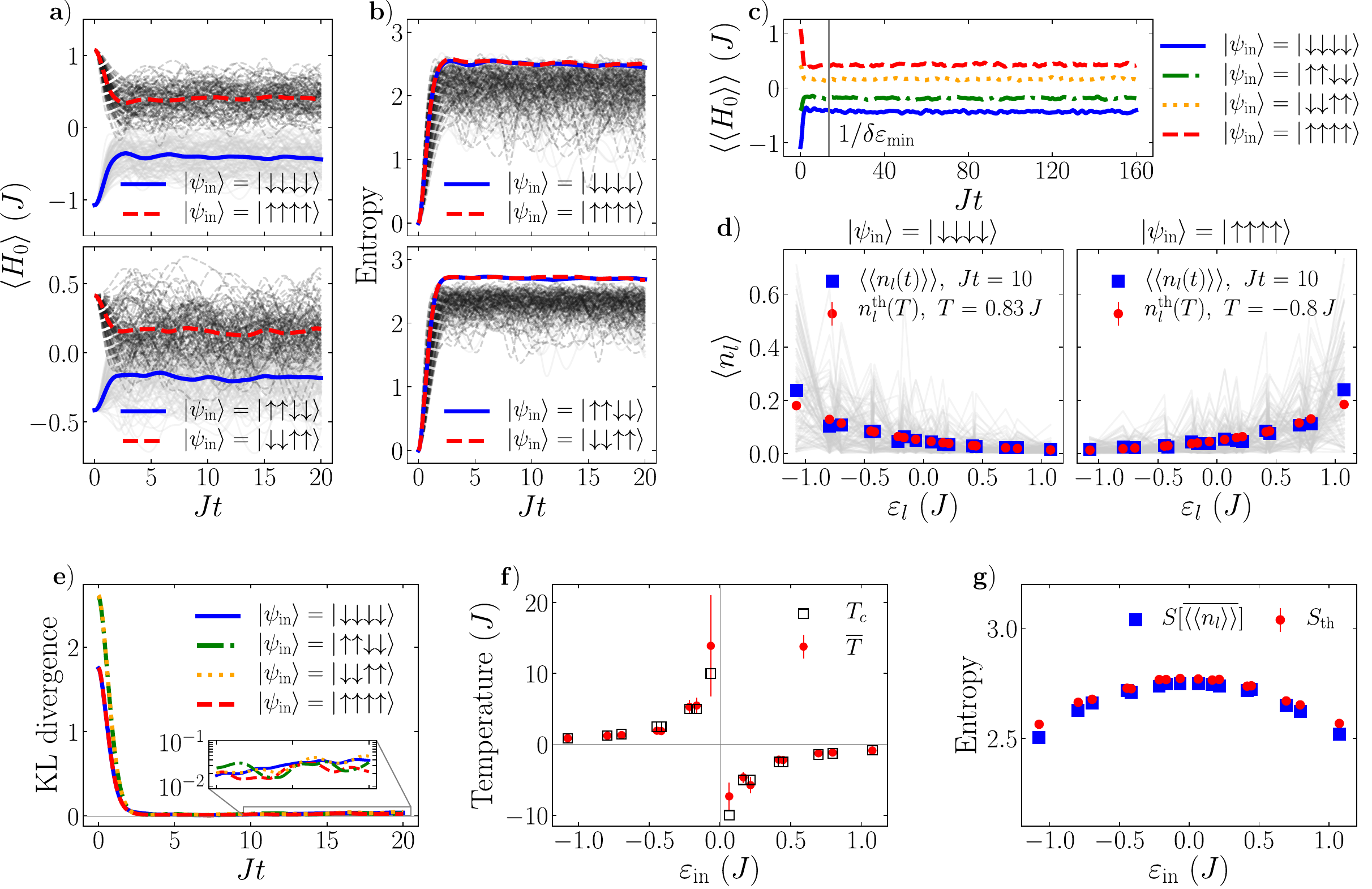} 
\caption{\small \label{fig:ED}  {\bf Exact dynamics} of a four-qubit system after the global quench performed over $100$ realizations of random couplings. The qubits' frequencies are $\omega_j = 0.28 J, \, 0.38 J,\, 0.63 J, \, 0.86 J$. {\bf a}) Expectation value of $H_0$ for four initial states as a function of time after the quench. The solid blue and dashed red curves show $\langle\langle H_0(t) \rangle\rangle$ for a given initial state, while the dim gray curves depict $\langle H_0(t) \rangle$ for each single realization of the quench protocol. {\bf b}) Diagonal entropy production evaluated with the average occupation probabilities (\ref{n_av}) for the same four initial states displayed by the solid blue and dashed red curves. The gray curves show the diagonal entropy, $ - \sum_l |\langle l| \psi(t)\rangle|^2  \ln |\langle l| \psi(t)\rangle|^2$, for individual realizations of the quench protocol. {\bf c}) Average expectation value $\langle \langle H_0(t) \rangle \rangle$ evolved at times beyond the inverse average minimal many-body level spacing of the Hamiltonian $H$, $\delta \varepsilon_{\rm min}=0.07 J$. {\bf d}) Occupation probabilities upon equilibration after quenching the qubits prepared in the ground state $|\psi_{\rm in } \rangle = |\!\downarrow\downarrow\downarrow\downarrow\rangle$ (left) and the highest excited state $|\psi_{\rm in}\rangle = |\!\uparrow\uparrow\uparrow\uparrow\rangle$ (right). The blue squares are the average occupations (\ref{n_av}) and the red dots show the thermal fit (\ref{n_th}). {\bf e})  Kullback-Leibler divergence between the average occupation probabilities (\ref{n_av}) and  the Gibbs distribution (\ref{n_th}) with temperature $T=\overline{T}(\varepsilon_{\rm in})$. The inset shows the late-time behavior of the KL divergence in a log scale. {\bf f}) Equilibrium temperature $\overline{T}(\varepsilon_{\rm in})$ as a function of the initial state's energy. The fitted temperature's error bar increases near the half-band as the system approaches the equally populated infinite temperature state. $T_c$ denoted by empty squares is the temperature determined from the canonical ensemble and total energy conservation. {\bf g}) Blue squares show the diagonal entropy evaluated with the mean occupation probabilities $\overline{\langle\langle n_l\rangle\rangle}$ averaged over the equilibrium time interval (from $t=2.5/J$ to $t=20/J$) for all the initial states of the system. Red dots display the thermal entropy, $S_{\rm th} = - {\rm Tr} \rho_{\rm th} (\overline{T}(\varepsilon_{\rm in})) \ln \rho_{\rm th}(\overline{T}(\varepsilon_{\rm in}))$. We determine the error bars for the thermal data displayed by the red dots from the covariance of the thermal fit of the average occupations shown by blue squares. The error bars for the average occupations stem from standard errors of the average over realizations of ${\cal V}$ and are within the size of the marker.
}
\end{figure*}

As shown in Fig.~\ref{fig:ED}{\bf a}, the average expectation value (\ref{H0_av}) becomes stationary at $t_{\rm eq} \approx 2.5/J$ up to small finite-size oscillations. On the contrary, the expectation value for each realization of random couplings persistently oscillates at high amplitude. The equilibration of the mean (\ref{H0_av}) appears along with the building up of the diagonal entropy, $S(t) = - \sum_l \langle \langle n_l(t) \rangle \rangle \ln \langle \langle n_l(t) \rangle\rangle$, which agrees with the properties of the thermodynamic entropy \cite{Polkovnikov2011} and approaches an average stationary value $S(t > t_{\rm eq})$ in Fig. ~\ref{fig:ED}{\bf b}. 
Alongside, in Figs.~\ref{fig:ED}{\bf a}, we note that the equilibrated observables depend on the choice of the initial state. To rule out the possibility that our simulation has reached a quasi-equilibrium steady state, such as prethermalization \cite{Berges2004Prethermalization, Langen2016}, rather than fully equilibrated, we explore the dynamics of the observables at times far exceeding $t_{\rm eq}$. Even though prethermalization is often associated with a weak breaking of integrability, it was recently observed in a trapped ion system with no conserved charges \cite{Neyenhuis2017Prethermalization}. In this case, achieving eventual thermalization and losing the information of the initial state requires evolution up to a timescale of the inverse minimal level spacing \cite{Gong2013Prethermalization}. In Fig.~{\ref{fig:ED}}{\bf c}, we demonstrate that the average expectation value (\ref{H0_av}) persists at times far exceeding the inverse minimal many-body level spacing of $H$, indicating complete equilibration, yet remains dependent on the initial state, which reflects the average total energy conservation. 

Having established equilibration (\ref{equilibration_2}), we explore the relaxation process further.
Since the the Hamiltonian $H=H_0 +{\cal V}$ is chaotic, it is natural to expect that the system equilibrates and thermalizes if it is large enough \cite{DAlessio2016}. For a small system of $N=4$ qubits, a dynamic observable averaged over an ensemble of realizations equilibrates after an interaction-dependent timescale $t_{\rm eq}$, even though an observable computed with an individual realization of the quench protocol may severely oscillate. Generally, one can express the evolved state of the system as $|\psi(t)\rangle = e^{-i (H_0 +{\cal V}) t}|\psi_{\rm in}\rangle =\sum_l c_l(t) |l\rangle$, where we specifically choose the basis of eigenstates of the initial Hamiltonian $H_0$. Accordingly, the coefficients of this representation $c_l(t) = \langle l| \psi(t)\rangle $ are related to the probability of occupying the $l$-th state at time $t$: $\langle n_l(t) \rangle=|c_l(t)|^2$. For $t>0$, these occupation probabilities begin to deviate from their initial values and become dynamically redistributed, evolving with the Hamiltonian (\ref{H}). If the system equilibrates, the probabilities to occupy the $l$-th state of $H_0$ should become stationary, describing an equilibrium probability distribution. Here we check if the average occupation probabilities become thermal.

Generically, describing the equilibrium state of a quantum many-body system requires a generalized Gibbs ensemble to account for the conserved quantities. The generalized Gibbs ensemble reduces to the conventional Gibbs ensemble if the conserved quantities are limited to total energy, particle number, and momentum \cite{Rigol2007Relaxation}. 
Since our quench protocol respects only the total energy conservation on average, we compare the averaged occupation probabilities upon equilibration with those predicted by the usual Gibbs distribution: 
\begin{equation}
    n_l^{\rm th}(T) = \frac{e^{- \varepsilon_l/T}}{Z}, \quad Z(T) = \sum\limits_{l=0}^{L-1} e^{-\varepsilon_l/T}. \label{n_th}
\end{equation}
By doing that, we aim to achieve thermal occupations for the initial system, assuming that the emergent temperature $T$ absorbs the effects of the interaction. At $t = 10/J > t_{\rm eq}$, we observe that while the occupation probabilities $\langle n_l(t)\rangle$ stay irregular for each individual realization of the quench protocol,
the average occupation probabilities (\ref{n_av}) are well-approximated with the occupation probabilities  $n_l^{\rm th}(T)$. To illustrate this agreement, we fit the average occupation probabilities (\ref{n_av}) with the thermal ones (\ref{n_th}) obeying the Gibbs distribution with the temperature $T$ being a fitting parameter. We compare $\langle\langle n_l(t) \rangle\rangle$ with the thermal fit $n_l^{\rm th}(T)$ in Fig.~\ref{fig:ED}{\bf d}.  

\begin{figure*}[t!!]
\center
\includegraphics[width=1.\linewidth]{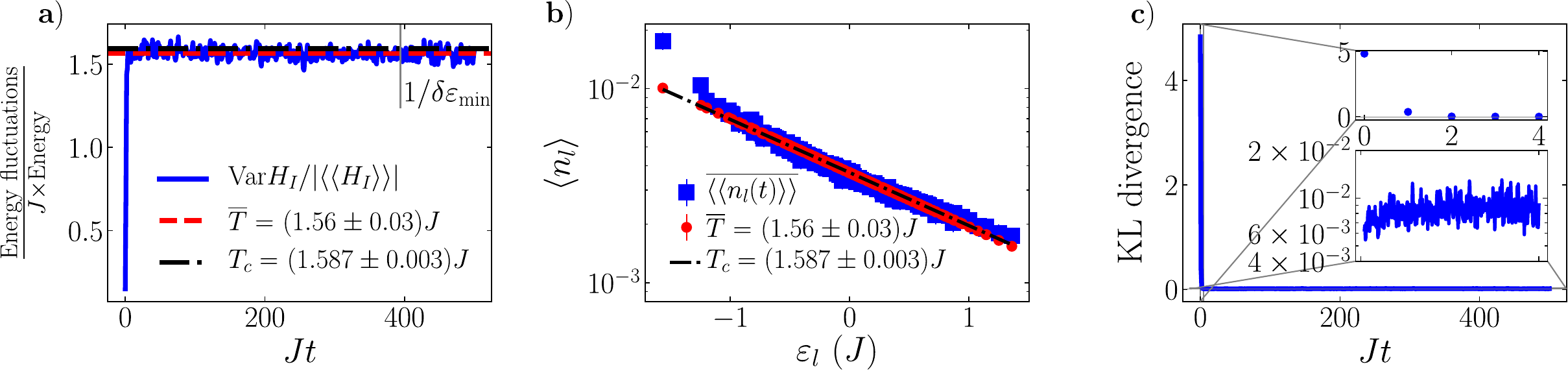}  
\caption{ \small  {\bf Thermalization dynamics of the non-integrable Ising model} with eight qubits and a periodic boundary condition. {\bf a}) Energy fluctuations-to-energy ratio of $H_I$ as a function of time after quenching ${\cal V}$. Horizontal lines show thermal predictions of the energy fluctuations-to-energy ratio, with temperatures determined from fitting the mean occupation probabilities averaged over time in the equilibrium regime ($\overline{T}$) and from total energy conservation ($T_c$). $\delta \epsilon_{\rm min} \approx 0.00254 J$ is the average minimal level spacing of the full Hamiltonian $H$. {\bf b})  Mean occupation probabilities for eigenstates of $H_I$, time-averaged over the equilibrium regime, in log scale. Blue squares are dynamical occupations averaged over 100 realizations of random couplings and time (from $t=10/J$ to $t=500/J$). The error bars are standard errors accumulated when averaging over realizations of ${\cal V}$, averaged over the equilibrium time interval. Red dots represent the thermal fit of the averaged dynamical data. The error bars are found from the square root of the covariance of the fitting parameters and are within the size of the marker. The dash-dotted black line shows thermal occupations with temperature determined by the total energy conservation (\ref{def_T_c}). {\bf c}) Kullback-Leibler (KL) divergence between  mean occupation probabilities (\ref{n_av}) and thermal occupations with equilibrium temperature $\overline{T}$. The top inset shows the first five time points of the KL divergence, illustrating that it approaches the equilibrium value in $t_{\rm eq} \sim {\cal O}(1)/J$. The bottom inset shows the KL divergence along the equilibrium time interval.  
\label{fig:Ising}
}
\end{figure*}

As expected upon thermalization, the occupations depend on a single parameter – temperature, which, in our case, is set by the energy of the initial state. One may adjust the temperature's absolute value by tuning the characteristic interaction strength, providing additional control over the thermalization process. The temperature retains the memory of the total energy persisting as a constraint throughout the relaxation, which makes the equilibrium observables initial-state-dependent. The choice of the initial state also defines the temperature's sign, giving rise to negative absolute temperatures.
Plotting the equilibrated mean occupation probabilities for the protocol started from the ground state (Fig.~\ref{fig:ED}{\bf d} {\it left}) and the highest excited state (Fig.~\ref{fig:ED}{\bf d} {\it right}) of the qubits, we find a population inversion in the second case, which implies $T<0$. While temperature is usually restricted to be non-negative for an unbounded energy spectrum to secure convergence of the partition function, negative absolute temperatures have been experimentally observed in systems with bounded spectra \cite{Purcell1951, Hakonen1994, Medley2011, Braun2013, Gauthier2019Giant, Johnstone2019Evolution, Nettersheim2022Power, Baudin2023, Spiecker2023}, as in our simulation. 

To justify the quantitative agreement between the mean occupation probabilities and the Gibbs distribution, we consider the Kullback-Leibler (KL) divergence \cite{Nielsen2009Quantum}. The KL divergence serves as a measure of distance between the two probability distributions. It is non-negative by definition and equals zero only if the two probability distributions coincide. Since the average observables still possess small finite-size oscillations, as seen in Fig.~\ref{fig:ED}{\bf a}, to find the equilibrium temperature $\overline{T}$, we additionally average the mean occupations over a fixed time interval in the equilibrium regime (from $t =  2.5 / J$ to $t =20 / J$) and approximate the result with the thermal distribution (\ref{n_th}) with $T=\overline{T}$. Then, we compare $n_{\rm th}(\overline{T})$ with the dynamic occupations (\ref{n_av}) via the KL divergence \cite{Gnezdilov2023, Ohanesjan2023}:
\begin{equation}
    D(  \langle  \langle n_l(t) \rangle  \rangle || n_l^{\rm th}(\overline{T}) ) \!=\! - S(t) \!+\! \frac{\langle\langle H_0(t) \rangle\rangle \!-\! F(\overline{T}) }{\overline{T}}, \label{KL}
\end{equation}
where $F(\overline{T}) \!=\! -\overline{T} \ln Z(\overline{T})$ is the thermodynamic free energy. For several initial states, we determine $\overline{T}$ and plot the KL divergence (\ref{KL}) as a function of time after the quench in Fig.~{\ref{fig:ED}}{\bf e}. The KL divergence decays for $t>t_{\rm eq}$ to the value $\sim 10^{-2}$ which quantifies the accuracy of the thermal occupation probabilities. 

Next, we evaluate the equilibrium temperature for $16$ initial states given by the eigenstates of $H_0$. Due to energy conservation, if the system thermalizes, one may estimate the equilibrium temperature by matching the initial energy and a canonical ensemble prediction for the total energy: 
\begin{equation}
 \varepsilon_{\rm in} = \langle {\rm Tr} \,  (e^{-H/T_c} \, H)/({\rm Tr} \, e^{-H/T_c})\rangle, \label{def_T_c}
\end{equation}
where the angular brackets denote averaging over realizations of random couplings and $H$ is the total Hamiltonian $H_0+{\cal V}$. Solving this equation gives the canonical prediction $T_c$ for the temperature as a function of the initial energy. As seen in Fig.~\ref{fig:ED}{\bf f}, the equilibrium temperature $\overline{T}$ determined by fitting the average occupation probabilities in the equilibrium regime agrees well with the canonical temperature $T_c$. Analyzing the equilibrium temperature $\overline{T}$ versus the energy of the initial state $\varepsilon_{\rm in}$ in Fig.~\ref{fig:ED}{\bf f}, we observe that the equilibrium temperature is increasing as the initial state's energy is increased up to the middle of the spectrum. A subsequent rise of the initial energy changes the equilibrium temperature from plus to minus infinity. Temperatures $\pm \infty$ are equivalent and correspond to the maximum value of the thermal entropy that follows the semicircle-like function of the initial state's energy, as shown in Fig.~\ref{fig:ED}{\bf g}. Increasing the energy above the half-band decreases the entropy, implying that the negative temperatures appear above $T = +\infty$ in the energy hierarchy \cite{Landau1980, Braun2013}. 

So far, we have achieved thermal occupations of the many-body energy states for the system described by the Hamiltonian $H_0$ of four non-interacting qubits, starting the protocol from an eigenstate of $H_0$. To highlight the generality of our protocol, we choose the $N= 8$ Ising model with a periodic boundary condition, $H_I = -\omega \sum_{j=1}^N \sigma_j^x  - h \sum_{j=1}^N \sigma_j^y  - g \sum_{j =1}^{N-1} \sigma_{j}^y \sigma_{j+1}^y -g \sigma_N^y \sigma_1^y$, as  the initial Hamiltonian. We tune  $H_I$ to the paramagnetic phase ( $\omega> h, g$: $\omega = 0.15J$, $h=0.05 J$, $g = 0.1 J$) and begin the evolution with a  $\sigma^y$-ordered state, $|\psi_{\rm in}\rangle = |\! \uparrow\uparrow\ldots \uparrow\rangle$, which is not an eigenstate 
of $H_I$. 
In Fig.~\ref{fig:Ising}{\bf a}, we plot the ratio of the energy fluctuations, ${\rm Var} H_I = \langle\langle H_I(t)^2 \rangle\rangle-\langle\langle H_I(t) \rangle\rangle^2$, to the mean energy $\langle\langle H_I(t) \rangle\rangle$ in the initial system, reminiscent of a fluctuation-dissipation relation for thermal systems, where we average over one hundred realizations of random couplings in the quenched interaction ${\cal V}$. The energy fluctuations-to-energy ratio equilibrates and remains stationary up to weak oscillations In Fig.~\ref{fig:Ising}{\bf b}, we show the average occupation probabilities (additionally time-averaged over the equilibrium regime from $t = 10/J$ to $t = 500/J$). As in the case of the non-interacting initial Hamiltonian, we fit the average occupation probabilities by the Gibbs state's prediction (\ref{n_th}) to determine the equilibrium temperature $\overline{T}$. We also determine the canonical temperature $T_c$ from the total energy conservation (\ref{def_T_c}). Both temperatures agree within the error bar, as seen in Fig.~\ref{fig:Ising}{\bf b}, where we plotted the Gibbs distributed occupations of $H_I$ with the temperature $T_c$ with a dash-dotted black curve on top of the averaged data (blue squares) and the thermal fit (red circles). The average occupation probabilities are well-described by the thermal ensemble except for the edges of the spectrum, expected for a finite-size system. However, the temperature found in both ways is sufficient to correctly predict the equilibrium value of the energy fluctuations-to-energy ratio for $H_I$, as seen in Fig.~\ref{fig:Ising}{\bf a}. We find the thermal values of the energy fluctuations-to-energy ratio from the thermal averages: $\langle H_I \rangle_{\rm th} = \sum_{l=1}^{L-1} n_l^{\rm th}(T)\varepsilon_l$ and ${\rm Var}_{\rm th}\, H_I = \sum_{l=1}^{L-1} n_l^{\rm th}(T)\varepsilon_l^2 -\langle H_I\rangle_{\rm th}^2$ with $T = \overline{T}$, $T_c$, where $\varepsilon_l$ are eigenenergies of $H_I$. Note that the spectrum of $H_I$ has multiple degeneracies. We show the corresponding thermal values by two horizontal lines in Fig.~\ref{fig:Ising}{\bf a}. The dash-dotted black line ($T=T_c$) lies right atop the dashed red line ($T=\overline{T}$). The thermal predictions agree with the equilibrated energy fluctuations-to-energy ratio obtained in the exact dynamics, highlighting that our protocol leads to the thermal behavior of energy fluctuations and energy of the initial system as expected from the fluctuation-dissipation theorem.
Finally, in Fig.~\ref{fig:Ising}{\bf c}, we show the KL divergence (\ref{KL}) between the mean dynamic occupations and Gibbs distribution with the equilibrium temperature $\overline{T}$. As seen in insets in Fig.~\ref{fig:Ising}{\bf c}, the KL divergence equilibrates to values $\sim 10^{-2}$ in time $\sim {\cal O}(1)/J$, qualitatively similar to our analysis for the non-interacting initial system. Despite increasing the system size, we still find that the system equilibrates in $t_{\rm eq} \sim {\cal O}(1)/J$.

Even for a small system, thermalization to a finite positive or negative temperature equilibrium occurs from the post-quench dynamics of a quantum system within the time scale $\sim {\cal O}(1)/J$. Suitability for a small system, controllable equilibration timescale, and a lack of a thermostat make our thermalization mechanism possible to test experimentally on NISQ computers like IBMQ. Indeed, as we demonstrate below, executing the evolution of the interacting four-qubit system, described by the Hamiltonian (\ref{H}), is viable on IBMQ with a moderate number of two-qubit gates, which is often the main limiting factor in studying many-body physics on NISQ devices. Achieving thermal observables relies on averaging the expectation values, which can be done classically from the output of every realization of the quench protocol, each running independently on a quantum computer.

{\bf Quantum simulation.}
We implement our dynamic thermalization protocol on an IBM Quantum computer (IBMQ) named \texttt{ibm\_hanoi} for $100$ realizations of the quench protocol, each corresponding to a quantum circuit. Encoding quantum dynamics into a quantum circuit implies applying a sequence of single- and two-qubit gates that approximate a given realization of the evolution operator $U(t) = e^{-i t (H_0+ {\cal V})}$ to the initial state $|\psi_{\rm in}\rangle$. Executing a quantum circuit on IBMQ is noise-susceptible, mostly due to the imperfect application of two-qubit gates, namely, CNOT gates that perform entangling operations between qubits in our simulation.
Hence, our strategy to mitigate the error accumulated due to noise is: first, decreasing the number of CNOT gates required to perform the simulation as much as possible; second, simplifying the noise's structure from the remaining CNOT gates and accounting for it in the simulation results. 

\begin{figure*}[t!!]
\center
\includegraphics[width=0.94\linewidth]{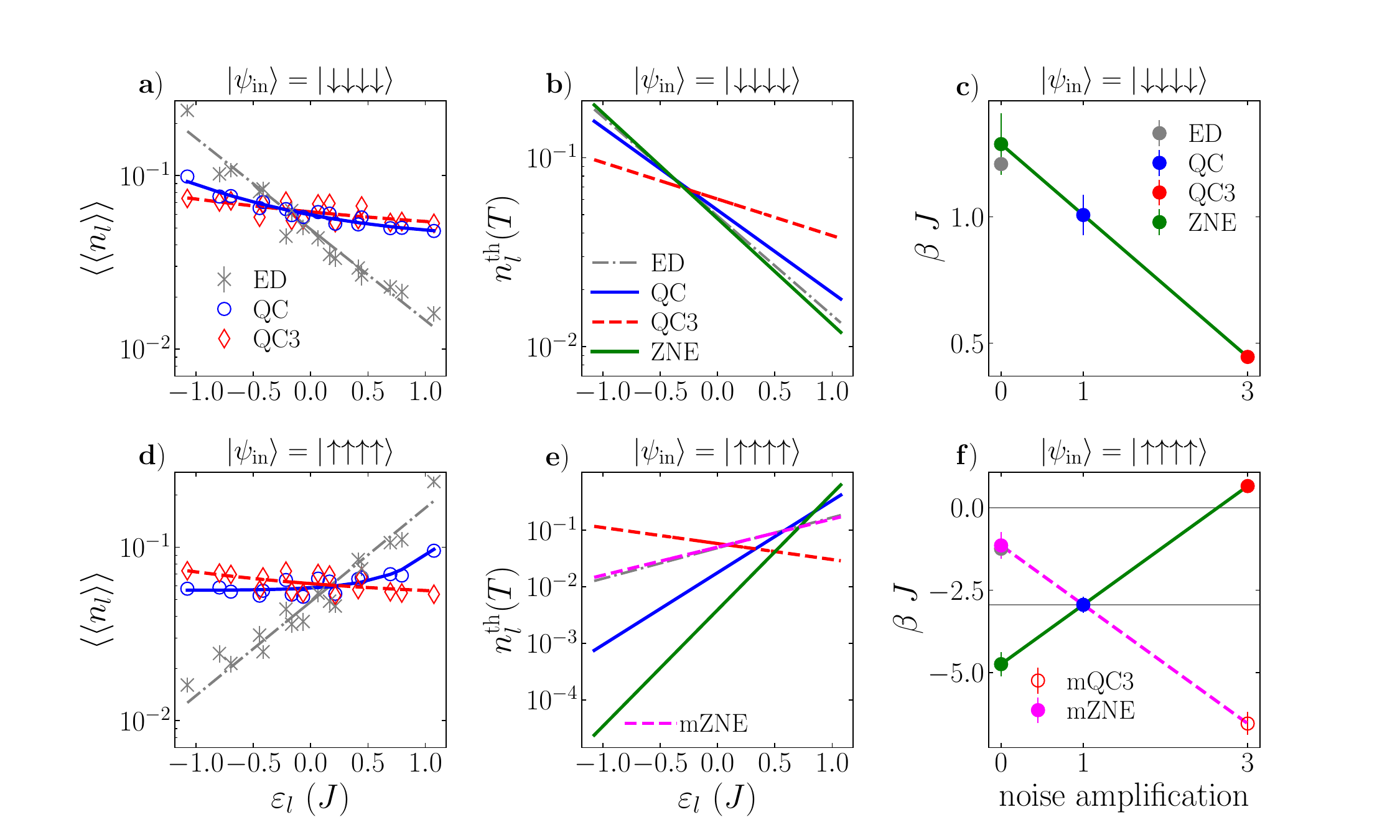} 
\caption{\small \label{fig:QS} {\bf Quantum simulation} results on the IBM Quantum computer. The evolution goes up to $t = 10/J$.  {\bf a}) Occupation probabilities averaged over $100$ realizations of the quench protocol with the initial state $|\psi_{\rm in}\rangle=|\! \downarrow\downarrow\downarrow\downarrow\rangle$: the gray crosses display the result of the exact dynamics (ED); the dash-dotted gray line is the thermal fit (\ref{n_th}) of the ED data; the blue circles show the outcome of the simulation on the quantum computer (QC) after error mitigation; the solid blue curve is the fit of the QC data with the model for noisy occupations (\ref{n_noisy}); the red diamonds and the dashed red curve show the same as QC but with tripled noise amplification (QC3).  {\bf b}) The Gibbs distribution for occupation probabilities (\ref{n_th}) with the temperatures determined from fitting the ED and quantum simulation data in the panel {\bf a}. The error bars for QC and QC3 data sets in panels {\bf a} and {\bf b} are within the size of the designated symbols. ZNE refers to the temperature value obtained with zero-noise extrapolation for $\beta = 1/T$ shown in the panel {\bf c}. Panels {\bf d}, {\bf e}, {\bf f} are analogous to {\bf a}, {\bf b}, {\bf c} but for the protocol starting with $|\psi_{\rm in}\rangle=|\! \uparrow\uparrow\uparrow\uparrow\rangle$. mQC3 is the mirror image of the QC3 value of the inverse temperature $\beta=1/T$ relative to the QC value of $\beta$, necessary for mirrored zero-noise extrapolation (mZNE) illustrated with the dashed magenta line in the panel {\bf f}. The frequencies of the qubits are the same as in Fig.~\ref{fig:ED}. The plots for occupation probabilities are in the log scale along the y-axis.
 The error bars for the occupation probabilities in ED arise due to the finite sampling of the $M=100$ realizations of the auxiliary interaction ${\cal V}$. 
We determine the error bars from the standard error: $\sigma_{\text{ED}}=\sigma_{\cal V}/\sqrt{M}$, where $\sigma_{\cal V}$ is the standard deviation of the distribution. For the QC data, error bars additionally account for the finite sampling of randomized compiled (RC) circuits and shots. In this case, they are given by:
$\sigma_{\text{QC}}=(\sigma^2_{\cal V}/M+\langle\sigma^2_{\text{RC+shot}}\rangle/M)^{1/2}$, where $\langle\ldots \rangle$ denotes the average over 
realizations of random couplings, and $\sigma^2_{\text{RC+shot}}$  is computed as described in Appendix E of Ref.~\cite{Perrin2023}. We find the error bars for inverse temperatures from the covariance of the thermal fit of the quantum data after averaging it over realizations of random couplings and applying preliminary error mitigation (such as RC and removing global depolarizing noise).  }
\end{figure*}

To fulfill the first task, we adapt variational circuit recompilation \cite{Heya2018, Khatri2019} to approximate the evolution operator $U(t)$ by a parametrized quantum circuit with $99.8\%$ accuracy, optimizing the tensor network decomposition of $U(t)$, before plugging it into the quantum computer (see Appendix \ref{app:VCR}). Using this method, we reduce the number of CNOT gates required for the simulation to $N_{\rm CNOT}=60$, discarding the need for Trotterization, which allows us to reach arbitrary times.

\begin{figure*}[t!!]
\center
\includegraphics[width=0.7\linewidth]{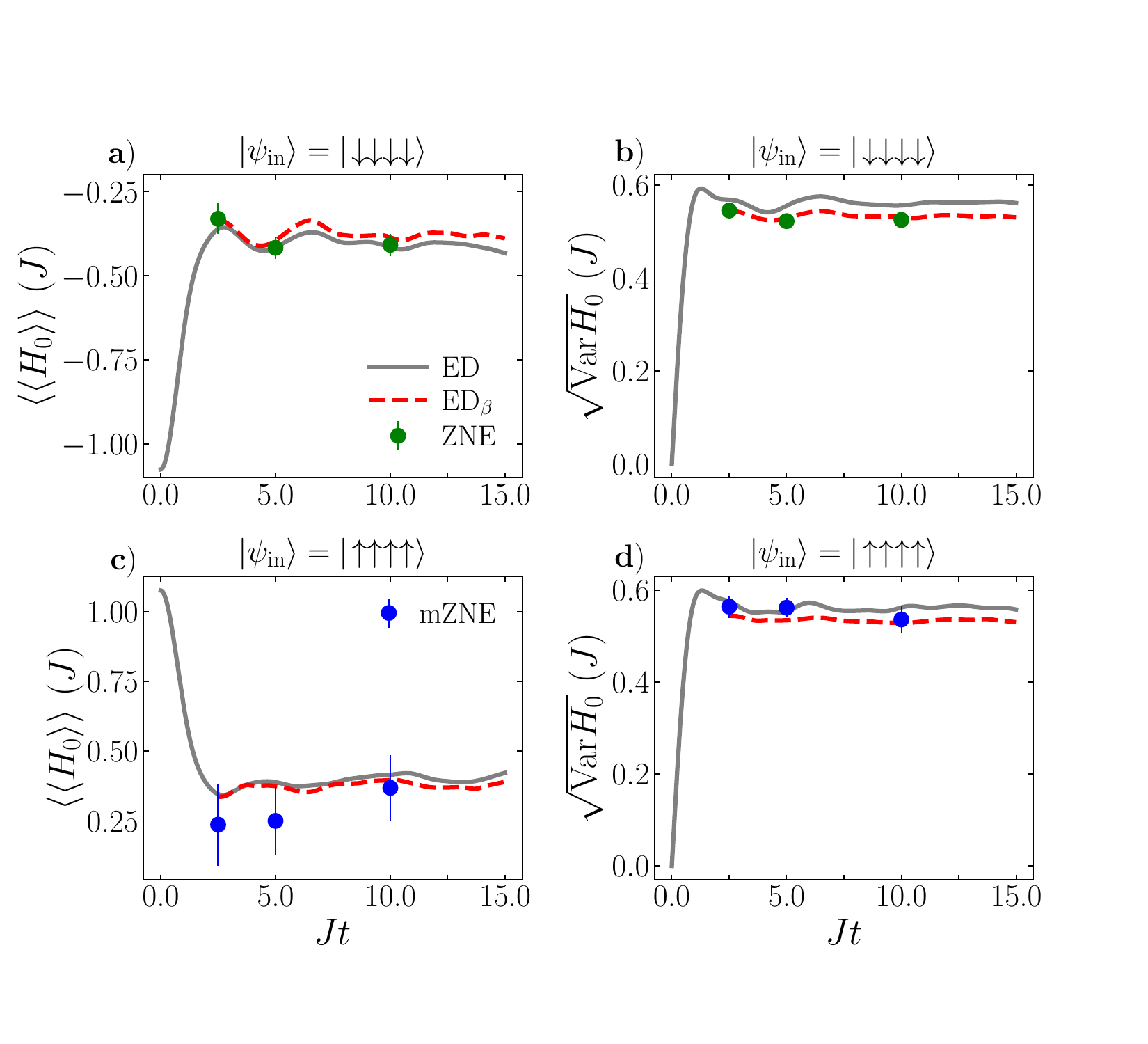} \vspace{-0.8cm}
\caption{\small \label{fig:QD} {\bf Exact vs Quantum dynamics} for the expectation value of initial Hamiltonian and its fluctuations. {\bf a}) Average expectation value of $H_0$ for $|\psi_{\rm in}\rangle = |\! \downarrow\downarrow\downarrow\downarrow\rangle$. ED denotes the exact dynamics, ED$_\beta$ approximates the ED curve with the thermal fit, and ZNE refers to the observables computed with thermal occupations obtained from the quantum computer in the zero-noise limit. {\bf b}) Fluctuations of $H_0$, $\sqrt{\langle\langle H_0^2(t) \rangle\rangle- \langle\langle H_0(t) \rangle\rangle^2}$, for $|\psi_{\rm in}\rangle = |\! \downarrow\downarrow\downarrow\downarrow\rangle$. Panels {\bf c} and {\bf d} show the same as in panels {\bf a} and {\bf b} but for the different initial state, $|\psi_{\rm in}\rangle = |\! \uparrow\uparrow\uparrow\uparrow\rangle$. mZNE designates the observables computed with the output of the mirrored ZNE approach we used to establish the zero-noise limit for states with negative temperatures. 
We compute the error bars for the quantum prediction of the mean energy from the uncertainty on the value of the inverse temperature $\beta_{\rm ZNE}$ (or $\beta_{\rm mZNE}$) found from the thermal fit in Fig.~\ref{fig:QS}: $\sigma_{H_0}=\sigma_{\text{(m)ZNE}}\sqrt{{\rm Var}_{\rm th}H_0}$, where ${\rm Var}_{\rm th}H_0 = \langle H_0^2 \rangle_{\rm th} - \langle H_0 \rangle_{\rm th}^2 = \sum_l n_l^{\rm th}(T)\epsilon_l^2 - (\sum_{l=0}^{L-1} n_l^{\rm th}(T)\epsilon_l)^2$ is the thermal energy variance at $T=1/\beta_{\text{(m)ZNE}}$ and $\epsilon_l$ are the eigenenergies of $H_0$.
Similarly, for the energy fluctuations, we use $\sigma_{\sqrt{{\rm Var}H_0}}=\sigma_{\text{(m)ZNE}}|\langle (H_0- \langle H_0\rangle_{\rm th})^3\rangle_{\rm th}|/(2\sqrt{{\rm Var}_{\rm th} H_0})$.  }
\end{figure*}

We address the second task using an extension of the randomized compiling (RC) approach \cite{Perrin2023} converting the noise, not only on qubits to which a CNOT gate is applied but also on the neighboring qubits, to an incoherent noise channel (see Appendix \ref{app:EM}). 
The incoherent noise contains local and global components, where the latter acts uniformly on every qubit in the circuit. To account for the global incoherent noise, we approximate the noisy quantum computer's output for average occupation probabilities (\ref{n_av}) upon equilibration as a weighted mixture of the average occupations unaffected by the global incoherent noise and a uniform distribution describing a maximally mixed (infinite temperature) state. Aiming to see if the observables evaluated on the quantum computer display thermal behavior, we chose the average occupations unaffected by the global incoherent noise in the form of the finite temperature Gibbs distribution (\ref{n_th}) so that the noisy average occupation probabilities read
\begin{equation}
    \langle\langle n_l(t>t_{\rm eq}) \rangle\rangle = f^{N_{\rm CNOT}} n_l^{\rm th}(T) + \frac{1-f^{N_{\rm CNOT}}}{2^N}. \label{n_noisy}
\end{equation}
The consistency of this choice settles by fitting the noisy quantum computer's data to the model (\ref{n_noisy}) where the temperature $T$ and the circuit's fidelity $f^{N_{\rm CNOT}}$ play a role of the fitting parameters. 

In Fig.~\ref{fig:QS}, we show the data for the average occupation probabilities evolved up to $t=10/J>t_{\rm eq}$ for two initial states $|\psi_{\rm in}\rangle = |\!\downarrow\downarrow\downarrow\downarrow\rangle, | \! \uparrow\uparrow\uparrow\uparrow\rangle$. For $|\psi_{\rm in}\rangle = $ $|\!\downarrow\downarrow\downarrow\downarrow\rangle$, in Fig.~\ref{fig:QS}{\bf a}, we notice that the average occupation probabilities obtained from the quantum computer (QC) deviate from the characteristic exponential shape of the Gibbs distribution fitted to the result for the exact dynamics (ED). However, the QC data is well-approximated by the two-parameters function (\ref{n_noisy}), as seen from the solid blue fitting curve. Extracting the temperature from fitting the QC data to the model (\ref{n_noisy}), we plot the corresponding thermal occupations (\ref{n_th}) in Fig.~\ref{fig:QS}{\bf b}. 

The temperature obtained from the quantum simulation is higher than the temperature from the thermal prediction of the exact dynamics, as we observe from the occupations' slopes in Fig.~\ref{fig:QS}{\bf b}. We attribute this heat-up to the part of the hardware noise so far unattended. Indeed, above, we have discussed the effect of the global incoherent noise, yet the local incoherent noise may be present in the system. We account for the remaining noise using the zero-noise extrapolation (ZNE) method \cite{Temme2017, Li2017, Endo2018, Dumitrescu2018, GiurgicaTiron2020, He2020}, which involves repeating the experiment with artificially amplified noise and extrapolating the measurement results to the noiseless limit. For incoherent noise, a linear change in noise amplification generally leads to exponential change in an observable \cite{Endo2018}. From this perspective, the inverse temperature $\beta=1/T$ in the exponent of thermal occupations in Eq.~(\ref{n_noisy}) may absorb the impact of the remaining noise. Elaborating on this observation, we repeat the experiment with the tripled noise by tripling every CNOT gate in the circuit. The average occupation probabilities in the quantum simulation with tripled noise amplification (QC3) remain well within the incoherent noise approximation, (\ref{n_noisy}) as shown in Fig.~\ref{fig:QS}{\bf a}. In turn, the temperature determined from the QC3 data (see Fig.~\ref{fig:QS}{\bf b}) increases further compared to the QC result, highlighting how the hardware noise on NISQ devices can lead to the temperature rise in the simulated observables. In Fig.~\ref{fig:QS}{\bf c}, we extrapolate the inverse temperature to the noiseless case using a linear fit based on two noise amplification values (QC, QC3). As the inverse temperature appears in the exponent of the noisy occupations (\ref{n_noisy}), our method effectively performs an exponential extrapolation on the observables after clearing the effect of the global incoherent noise from the quantum data. As a result, the value of $\beta_{\rm ZNE} = (1.29 \pm 0.12)/J$ determined from ZNE lies within the error bar with $\beta_{\rm ED} = (1.21 \pm  0.04)/J$ found from fitting the occupations computed in the exact dynamics with the Gibbs distribution (\ref{n_th}).  The canonical prediction for $\beta_c = 1/T_c = 1.20/J$ based on energy conservation well agrees with exact dynamics and experimental value of the inverse temperature upon error mitigation.

The state evolved from the highest energy state, $|\psi_{\rm in}\rangle = | \! \uparrow\uparrow\uparrow\uparrow\rangle$, equilibrates to the inversely populated state in the quantum simulation as it does for the exact dynamics, with the QC results well-described by the noisy occupation probabilities (\ref{n_noisy}) with $T<0$, as shown in  Figs.~\ref{fig:QS}{\bf d},{\bf e}. Tripling the noise on the quantum computer leads to the system's equilibration to a state with $T>0$. Indeed, negative temperature states are unstable when interacting with an environment \cite{Landau1980}, which makes them particularly challenging to prepare. Since negative absolute temperatures are hotter than $T=+\infty$, with $T=0^-$ being the hottest \cite{Landau1980, Braun2013}, from a thermodynamic standpoint, the environment (considered at $T>0$) cools the system at $T<0$ to a positive temperature. Thinking of the hardware noise as a consequence of a quantum computer being an open system, one expects the inverse populations to be unstable against noise, as we observe for the QC3 occupation probabilities in Figs.~\ref{fig:QS}{\bf d},{\bf e}. However, in our case, the noise plays a substantial role in the dynamic formation of the system's equilibrium state, being determined by the hardware and the circuit layout rather than describing an independent thermal entity interacting with the system equilibrated to a negative absolute temperature. Hence, it is hard to predict a threshold noise preserving $T<0$ equilibrium. 
Therefore, performing zero-noise extrapolation to the noiseless negative temperature case requires caution. Upon tripling the noise, the temperature crosses infinity ($\beta = 0$) and has a high positive value (small positive $\beta$), implying cooling the system. Thereby, ZNE for the QC and QC3 inverse temperatures predicts the value of $\beta_{\rm ZNE} = (-4.74\pm  0.37)/J$ quite distant from the ED result, $\beta_{\rm ED} = (-1.25 \pm 0.04)/J$, in Fig.~\ref{fig:QS}{\bf f}. To extrapolate to the regime of a closed system, we take a heuristic approach and assume that the increasing noise heats the system monotonically, restricting $T$ to the $T<0$ sector and preventing relaxation to $T>0$ equilibrium. In this way, we account for $\Delta\beta = |\beta_{\rm QC}-\beta_{\rm QC3}|$ and mirror the QC3 data point relative to the QC accordingly. Then we apply ZNE for the mirrored QC3 (mQC3) and QC points, shown in the dashed magenta line in Fig.~\ref{fig:QS}{\bf f}. The value of $\beta_{\rm mZNE} = (- 1.15 \pm  0.41)/J$ determined with this mirrored ZNE (mZNE) method is within the error bar from the result of the exact dynamics.

Having obtained the average thermal occupation probabilities in the zero-noise limit, we check if the quantum computer predicts thermal values of the dynamic observables in the equilibrium regime. We repeat the protocol for times $t= \lbrace 2.5, 5, 10\rbrace/J$ and compare the IBMQ results with the exact dynamics and the thermal approximation of ED (ED$_\beta$) for the average expectation value $\langle\langle H_0(t) \rangle\rangle$ and its fluctuations $\sqrt{{\rm Var}H_0} = \sqrt{\langle\langle H_0(t)^2 \rangle\rangle- \langle\langle H_0(t)\rangle\rangle^2}$. For the quench protocol started with the ground state of $H_0$, $|\psi_{\rm in}\rangle=|\! \downarrow\downarrow\downarrow\downarrow\rangle$, the mean expectation value (\ref{H0_av}) computed with the thermal occupations predicted by the quantum computer agrees with the ED and  ED$_\beta$ results within the error bars, as seen in Fig.~\ref{fig:QD}{\bf a}. As for the fluctuations of $H_0$ shown in Fig.~\ref{fig:QD}{\bf b}, the quantum computer's effective output lies right on the thermal curve, indicating that the fluctuations computed on a quantum computer are closer to the thermal equilibrium value than the prediction of the exact dynamics. For $|\psi_{\rm in}\rangle=|\! \uparrow\uparrow\uparrow\uparrow\rangle$, the quantum computer predicts negative temperatures upon equilibration. In Figs.~\ref{fig:QD}{\bf c},{\bf d}, we demonstrate that the IBMQ output provides for $\langle\langle H_0 \rangle\rangle$ and $\sqrt{{\rm Var}H_0}$ within the error bars from both ED and ED$_\beta$ predictions for $T<0$ regime. 

{\bf Discussion.}
We show how thermal observables can dynamically emerge in a closed quantum system as a result of averaging over repeated instances of a global quench initializing Gaussian random non-local interactions between the system's subparts.  After the quench, the system pursues evolution governed by the chaotic Hamiltonian falling into the class of Gaussian unitary ensemble of random matrices. Averaging the observables over many realizations of the quench protocol leads to their equilibration (\ref{equilibration_2}) and thermalization (\ref{thermalization_2}) in time $t_{\rm eq} \sim {\cal O}(1)/J$, where $J$ is the characteristic interaction strength. The resulting averaged probabilities of occupying eigenstates of the initial system in the equilibrium regime, $t>t_{\rm eq}$, follow the Gibbs distribution with positive or negative absolute temperatures depending on the initial state's energy, conserved on average in the course of after-the-quench dynamics. One can additionally tune the absolute value of equilibrium temperature by adjusting the interaction strength $J$. This dynamic thermalization is particularly convenient to examine on NISQ devices such as IBMQ since it does not rely on the long-time evolution, large system size, or auxiliary qubits to emulate a thermostat but solely on the repetition of the quench protocol with different sets of random coupling constants.  

Adapting the established error-mitigation techniques, we implement our thermalization protocol on the IBMQ machine using a circuit depth of $60$ CNOT gates. We obtain thermal occupations for the sixteen many-body energy states of the four-qubit system, populated according to positive or negative absolute temperatures depending on the choice of the initial state. Respectively, the observables and their fluctuations found in quantum simulation agree with the thermal predictions. This utility of the digital quantum simulator for producing thermal observables may be attributed to the notion of the quantum computer as an open system and our implementation of the thermalization protocol. First, we average the expectation values over realizations of the quench protocol aiming at mimicking the thermodynamic limit and facilitating thermalization in a relatively small system.
One way to think of thermalization in a small system is to consider it as a subpart of a much bigger system in the thermodynamic limit, with the rest of the system acting as a thermal bath. A recent observation notes that averaging the evolution of a system with randomized interaction results in effective dissipative dynamics \cite{Bandyopadhyay2023}. Another recent study shows that adding a disorder in a simulated system in analog quantum simulations substantially reduces the necessary size of a bath for infinite temperature  deep thermalization \cite{Mok2024Optimal}.  In our case, the finite temperature in the equilibrated observables relies on the property of the auxiliary interaction that respects the system's total energy conservation on average in the quench protocol. In our experiment, we start with a pure state and observe dynamical relaxation to a finite temperature equilibrium for averaged observables without involving a thermal bath, with the mitigated noise seemingly falling into the same effectively dissipative channel and raising the temperature of the simulated observables.
Second, we apply zero-noise extrapolation that resembles taking a dissipationless limit in an open quantum system. Genuinely, these two limits are non-commutative, and their correct order, naturally implemented in our protocol, may be essential to justify the occurrence of the fluctuation-dissipation theorem in a quantum system \cite{Pastawski2007}.  Exploring this issue numerically for the Ising model as the initial Hamiltonian in our protocol, we find that the energy fluctuations-to-energy ratio equilibrates to the value predicted by temperature determined from the total energy conservation, resembling a fluctuation-dissipation relation for thermal systems. Alongside, in our simulation, the time for the observables to equilibrate is ${\cal O}(1)/J$, similar to the timescale of irreversible dynamics $T_3$ in an open quantum system, after which an unconstrained system thermalizes \cite{Sanchez2020}.

The advantage of the considered quench protocol is its inherently dynamic nature, which allows studying the occurrence of thermalization or, potentially, ergodicity breaking in quantum systems in real-time, efficiently engaging the limited resources of NISQ processors. We apply our quench protocol to study the emergence of the negative absolute temperature equilibrium on a digital quantum platform. While absolute temperatures are usually restricted to positive values due to kinetic energy, equilibrium at negative absolute temperatures is possible under certain conditions, in systems with a bounded energy spectrum. Bringing negative absolute temperature equilibrium states to reality inspired significant experimental efforts due to their non-trivial thermodynamic properties, e.g., losing entropy when absorbing energy.  By now, negative temperature states were experimentally observed in a few physical systems such as nuclear spins \cite{Purcell1951, Hakonen1994}, cold atoms \cite{Medley2011, Braun2013, Nettersheim2022Power}, two-dimensional quantum superfluids \cite{Gauthier2019Giant, Johnstone2019Evolution}, optical fibers \cite{Baudin2023}, and a two-level system environment in superconducting qubits \cite{Spiecker2023}. Here, we present an equilibrium state with negative absolute temperature arising from the relaxation process in a quantum system realized digitally on a quantum computer. 

{\bf Data Availability.}
The raw and processed quantum simulation data is available on Zenodo open repository \cite{Perrin2025Zenodo}.

{\bf Acknowledgements.}
We thank Alexander Balatsky, Igor Gornyi, Archana Kamal, Alexander Mirlin, Horacio Pastawski, Anatoli Polkovnikov, and Alexander Shnirman for stimulating discussions. HP acknowledges support from the Horizon Europe program HORIZON-CL4-2021-DIGITAL-EMERGING-01-30 via the project 101070144 (EuRyQa). HP and TS acknowledge funding from the state of Baden-Württemberg through the Kompetenzzentrum Quantum Computing Project QC4BW. AP was supported by the German Ministry of Education and Research (BMBF) within the project QSolid (FKZ: 13N16151). NG was supported by the Department of Physics and Astronomy, Dartmouth College.

\begin{appendix}

\begin{figure*}[t!!]
\center
\includegraphics[width=0.6\linewidth]{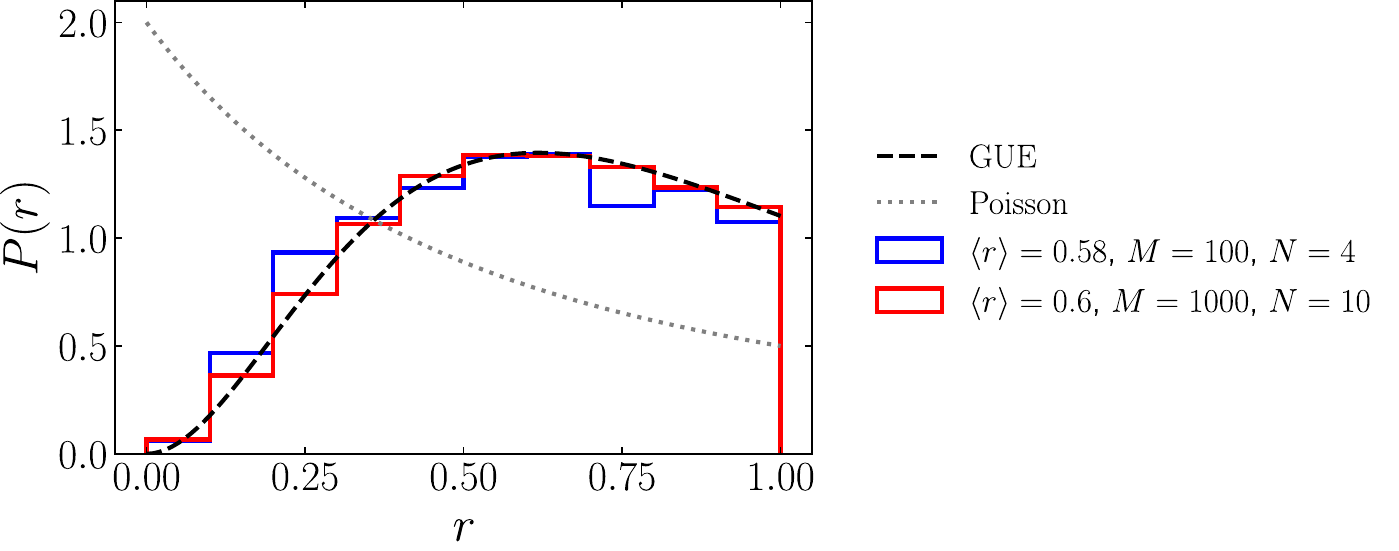}  
\caption{\small {\bf Probability distribution of the ratio of consecutive level spacings} for the Hamiltonian (\ref{H}). The blue histogram shows the probability distribution for a system of $N=4$ qubits with the same $M=100$ realizations we use for exact and quantum dynamics. We display the data averaged over $12$ ratios of consequential level spacings without two pairs of states at the edges of the spectrum. The red histogram is done for the system with $N=10$ qubits over $M=1000$ realizations, averaging the data over $\sim 10\%$ out of $2^{10}$ of many-body energy levels. For comparison, the black dashed line shows the  Gaussian unitary ensemble (GUE) predicted level statistics. The gray dotted line displays Poisson (integrable) behavior put for a visual reference. Both ($N=4$, $M=100$) and ($N=10$, $M=1000$) datasets saturate the GUE curve and possess the GUE characteristic value of the mean adjacent gap ratio $\langle r\rangle_{\rm GUE} \approx 0.6$.
\label{fig:WD}
}
\bigbreak
\center
\includegraphics[width=1.\linewidth]{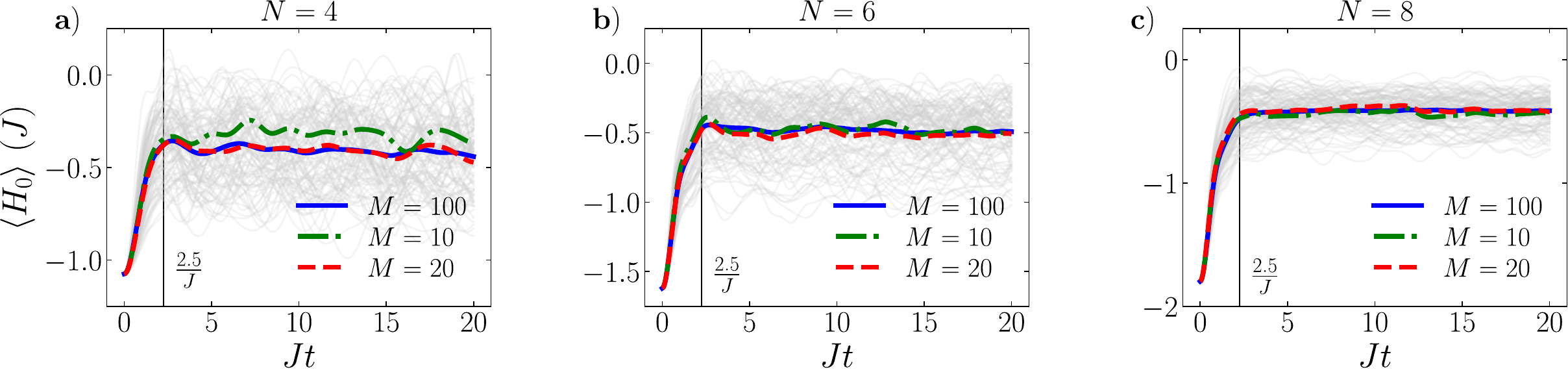}  
\caption{ \small   {\bf Equilibration of the multi-qubit system} in the initial state $|\psi_{\rm in}\rangle = |\downarrow\downarrow \ldots \downarrow \rangle$ after the quench. We plot the expectation value of $H_0$ as a function of time for individual realizations of random couplings (gray curves), averaged over $M=100$ realizations (solid blue curve), averaged over $M=10$ realizations (dash-dotted green curve), and averaged over $M=20$ realizations (dashed red curve) for different system sizes: {\bf a}) for $N=4$ qubits, {\bf b}) for $N=6$ qubits, and {\bf c}) $N=8$ qubits. The vertical black line marks the equilibration timescale at $t=2.5/J$. 
\label{fig:H0_NM}
}
\end{figure*}

\section{Choice of thermalization protocol}\label{app:RMT}

The choice of the interaction is essential for our protocol so that there would be no constraints preventing thermalization after the quench. We began with a ``constrained'' non-interacting Hamiltonian $H_0$ that can not be the sole source of thermalization dynamics. In our case, thermalization occurs due to the quenched random interaction ${\cal V}$ (with zero mean strength) between the qubits. Owing to the zero mean strength interaction, adding the latter to the initial Hamiltonian conserves only the total energy equal to the initial energy of the system on average. The inter-qubit interaction (\ref{V}) maps onto the random free fermions, also known as the SYK$_2$ model, an explicitly non-interacting theory. However, the combined Hamiltonian $H=H_0 + {\cal V}$ is non-integrable, as may be seen from its level statistics. 

The notion of energy level statistics allows characterizing integrability, localization, and chaos in quantum many-body systems. Chaotic behavior, closely related to thermalization \cite{DAlessio2016}, manifests in statistical properties of a system's many-body energy levels, falling in the Wigner-Dyson statistics described by a Gaussian orthogonal (GOE), unitary (GUE), or symplectic (GSE) ensemble of random matrices \cite{Bohigas1984}. On the contrary, integrability and localization are associated with level statistics following a Poisson distribution \cite{Berry1977}. Accordingly, changing disorder in a system may lead to a localization transition from Wigner-Dyson to Poisson level statistics \cite{Oganesyan2007Localization, Pal2010, Atas2013Distribution, Luitz2015, Giraud2022, Scoquart2024}. 

To quantify the level statistics in our model, we use the probability distribution for a ratio $0 \leq r_l \leq 1$  determined from two consecutive level spacings in a given part of a many-body spectrum \cite{Oganesyan2007Localization}: $r_l = {\rm min}\lbrace \delta_{l+1}, \delta_l \rbrace/ {\rm max}\lbrace \delta_{l+1}, \delta_l \rbrace$, where $\delta_l = \tilde{\varepsilon}_{l+1}-\tilde{\varepsilon}_l$. Note that here $\tilde{\varepsilon}_l$ are the eigenvalues of $H$. For the adjacent gap ratio $r_l$ following Wigner-Dyson statistics, the probability distribution is $P(r)= 2(1+r)^\beta/(1+r+r^2)^{1+\frac{3\beta}{2}}/Z_\beta$ with $Z_\beta$ being a normalization constant and $\beta$ is a symmetry index (do not confuse with an inverse temperature) equal to $1, 2$, and $4$ for GOE, GUE, and GSE, predicting mean values of the gap ratio $\langle r\rangle_{\rm GOE} \approx 0.54$, $\langle r\rangle_{\rm GUE} \approx 0.6$, and $\langle r\rangle_{\rm GSE} \approx 0.68$ \cite{Atas2013Distribution}. In the integrable case, for the Poisson spectrum, the distribution of the gap ratio is $P(r)=2/(1+r)^2$ with $\langle r \rangle_{\rm Poisson} \approx 0.39$ \cite{Oganesyan2007Localization}. Following earlier studies \cite{Oganesyan2007Localization, Pal2010, Atas2013Distribution, Luitz2015, Giraud2022, Scoquart2024}, we consider our total Hamiltonian $H=H_0 + {\cal V}$ in the regime when the system is large enough ($N=10$ with $2^N$ many-body energy levels) and plot the histogram for the gap ratio distribution for $M=1000$ disorder realizations averaged over $\sim 10\%$ of levels in Fig.~\ref{fig:WD}. The distribution function is in complete agreement with GUE prediction and reproduces the mean adjacent gap ratio $\langle r\rangle_{\rm GUE} \approx 0.6$. Reducing the system's size and the number of realizations to what we use in our experiment and choosing the parameters accordingly, we observe that the level statistics of the model remain well-described by GUE in Fig.~\ref{fig:WD}, highlighting an absence of integrability or localization.

Having established that the model is chaotic, i.e., described by a random matrix theory, one expects the observables evolved with the Hamiltonian $H$ to equilibrate and thermalize. However, one realization of random couplings may not be enough for equilibration. In Fig.~\ref{fig:H0_NM}, we show the after-the-quench equilibration process for $\langle H_0(t)\rangle$ for different system sizes ($N=4, 6, 8$), averaging the expectation values with several numbers of realizations $M=10, 20, 100$. Increasing the number of realizations leads to less finite-size oscillations in the average observable in the equilibrium regime.
The larger the system, the fewer disorder realizations are needed to suppress the remaining oscillations in the observable. 

Equilibration in the system results in thermal observables. In particular, we use the probabilities to occupy the many-body energy eigenstates of the initial Hamiltonian $H_0$. After the quench, the average occupation probabilities relax to a finite-temperature Gibbs distribution. This protocol allows for dynamically obtaining equilibrium observables at finite temperatures for a given system on a digital quantum processor without simulating a thermal bath. 

\begin{figure}[t!!]
\center
\includegraphics[width=1.\columnwidth]{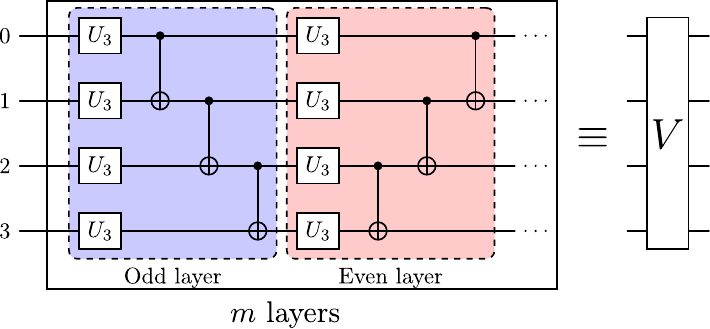}   
\caption{\small {\bf Quantum circuit ansatz} used for the variational recompilation of the circuit. A 4-qubit unitary $V$ is implemented as a parametrized $m$-layer quantum circuit of alternating odd and even layers. Each universal single-qubit gate $U_3$ is parametrized by $3$ distinct parameters, for a total of $n_p=12m$ parameters for $V$. Using our recompilation algorithm, we optimize the circuit over the set of $n_p$ parameters to minimize the trace distance between $V$ and the target unitary $U(t) = e^{-iHt}$. In our experiments, we use recompiled circuits with $m=20$ layers for a total of $60$ CNOT gates.
\label{fig:ansatz_recompilation}
}
\end{figure}

\section{Variational circuit recompilation}\label{app:VCR}
On IBMQ devices, errors largely originate from the application of noisy CNOT gates associated with an error rate of $\sim 1\%$. A naive compilation of the unitary operator $U(t)$  into a quantum circuit using, e.g., Qiskit's transpile function \cite{qiskit2024} on a linear qubit layout yields a quantum circuit with as much as  $238$ CNOT gates, which would result in a global error rate close to $100\%$. This number of entangling gates originates from the all-to-all interaction (\ref{V}) essential for our thermalization mechanism. To minimize the number of noisy CNOT gates required for the simulation, we take an alternative approach to recompile our quantum circuits, adapting a variational approach similar to the one presented in Refs.~\cite{Heya2018, Khatri2019}, which has already shown promising results on NISQ devices \cite{Chen2023, Koh2022, Koh2022b}. The general idea of the method is to consider an ansatz quantum circuit with a given CNOT gate structure and parametrized single-qubit gates and optimize these parameters until the circuit implements a suitable approximation of the target unitary $U(t)$. For few-qubits circuits, the optimization is done on a classical computer using a tensor network decomposition of both the ansatz and target circuit.

To perform the recompilation we first generate a parametrized ansatz circuit made of $m$ alternating layers of CNOT and universal $U_3$ single-qubit gates on every qubit as shown in Fig.~\ref{fig:ansatz_recompilation}. The universal single qubit gate $U_3$ gate is characterized by $3$ independent parameters, yielding $3 \times 4 = 12$ independent parameters per layer. Thus, for $N=4$ qubits, the ansatz circuit of $m$ layers produces a 4-qubit unitary $V$, a function of $n_p = 12m$ parameters. Next, we use the Quimb python library \cite{Gray2018} to decompose both $V$ and our target unitary $U(t)$ into rank-2 tensor networks and compute the cost function $C(U,V) = 1- {\rm Tr} \{V^\dagger U\}$. Finally, we perform the optimization for $C$ using the limited-memory Broyden-Fletcher-Goldfarb-Shanno algorithm with box constraints (L-BFGS) \cite{Malouf2002, Andrew2007} in combination with a basin-hopping method \cite{Li1987, Wales1997, Wales1999}. This combination is particularly suitable for optimization problems with many parameters and many local minima. Using this method, we recompile our evolution operator $U(t)$ for each realization of the quench protocol into quantum circuits structured as in Fig.~\ref{fig:ansatz_recompilation} with $m=20$ layers i.e. $60$ CNOT gates with an average fidelity of the order of $1-C(U, V) = 99.8\%$. 

The optimization runs on a regular laptop in a few minutes. The advantages of this optimization method are a drastic reduction in the number of required CNOT gates by $178$ and no need for a Trotter approximation, which allows us to run the experiment up to arbitrary long times with good precision, crucial to studying thermalization.

\section{Error mitigation}\label{app:EM}
Simulating the time evolution of a quantum-mechanical system on noisy NISQ devices, like IBMQ quantum computers, is susceptible to noise, generating output errors due to multiple sources. On IBMQ devices, it is known that errors stem principally from imperfect measurements, and from the application of imperfect two-qubit entangling gates, in our case, CNOT gates. 

{\bf Measurement errors.}
To mitigate measurement errors, we execute every circuit with a sampling of $1000$ shots and correct the outcome using the readout error correction technique that involves matrix inversion through the iterative Bayesian unfolding method \cite{Nachman2020}.

{\bf Randomized compiling.}
To mitigate errors arising from the two-qubit gates application, we aim to simplify the noise structure, which allows for the implementation of efficient error mitigation techniques. To do so, we follow the randomized compiling (RC) approach \cite{Kern2005, Wallman2016, Hashim2021} which enables one to transform any coherent noise affecting CNOT gates into an effective incoherent Pauli noise channel \cite{Cai2019}. RC entails generating alternative versions of the quantum circuit of interest, in which every CNOT gate is dressed with Pauli gates acting on individual qubits and drawn randomly from a set of Pauli matrices. These Pauli matrices are drawn in such a way that the dressed CNOT gates are still logically equivalent to an isolated CNOT gate, so that every alternative version of the circuit still implements the same unitary operation. The measured expectation values stemming from each randomly compiled circuit are then averaged, yielding the expectation value that one would have obtained if all CNOT gates were only affected by a purely incoherent noise channel.  

In addition to the qubits on which the CNOT gate is applied, a noisy CNOT gate may also lead to coherent errors on neighboring qubits. In our case, this can be formalized by extending the noise channel associated with a given CNOT gate to all qubits in the register. This noise, known as spillover crosstalk noise \cite{Sarovar2020}, can also be made purely incoherent by extending the RC method to neighboring qubits. In addition, since a priori no logical operation is being applied on these other qubits, the RC procedure can be improved by applying extra random $\pi/2$ rotations on top of the Pauli matrices, as prescribed in \cite{Perrin2023}. One can show that by doing this, the effective incoherent noise channel obtained after averaging over crosstalk RC circuits is a combination of a local Pauli noise channel for the active qubits of the CNOT gates, a local depolarizing noise channel acting on the neighboring qubits, and a global depolarizing noise channel acting uniformly on all qubits. In our quantum simulation, we run $100$ crosstalk RC circuits for every realization of the quench protocol, and average the read-out observables, effectively realizing the noise channel described above. We have checked for this number of circuits that the crosstalk RC averages has clearly converged to its infinite sampling value.
Using our knowledge of the structure of the effective incoherent noise channel acting on our qubits, we now implement an error mitigation strategy. 

{\bf Removing global depolarizing noise.}
After performing crosstalk RC, we know that at least a part of the effective noise acting on our circuit takes the form of an $N$-qubit global depolarizing noise channel. Due to the simplicity of this noise channel, its effects can be mitigated efficiently with a simple fit, as we show here. For a single CNOT gate, the $N$-qubit global depolarizing noise channel reads \cite{Nielsen2009Quantum}:
\begin{equation}
    \mathcal{E}_N(\rho)=f \rho +(1-f)\frac{\mathbb{1}}{2^N},
     \label{eq:depol_noise}
\end{equation}
where $\rho$ the density matrix, $f$ is the fidelity of a CNOT gate and $\mathbb{1}$ the identity matrix of size $2^N\!$. The rightmost term, which models the output of a global depolarizing error occurring, is the maximally mixed state with $N$ qubits. If such an error occurs on any CNOT gate during the execution of the circuit, then the qubits will remain in this state, as it is not affected by the action of subsequent unitary operations. Thus, it is straightforward to generalize Eq.~\eqref{eq:depol_noise} to the case of a circuit with $N_\text{CNOT}$ CNOT gates by replacing the fidelity of a single CNOT gate $f$ with a circuit fidelity $f^{N_{\rm CNOT}}$. Further averaging the quantum simulation data for occupation probabilities over $100$ realizations of the quench protocol, we interpret the noisy quantum computer's output within the model (\ref{eq:depol_noise}), leading to the expression (\ref{n_noisy}) for noisy average occupations. Then, we leverage the errors due to the local Pauli and depolarizing noise channels using zero-noise extrapolation (ZNE).

\end{appendix}

\bibliography{refs}

\end{document}